# Defect detection and NDE of low-modulus PMMA material using mechanical loading and WFT analysis

Jitendra Dhanotia[*], Amit Chatterjee[*], Vimal Bhatia[*], and Shashi Prakash[#]

[*] *Signals and Software Group, Discipline of Electrical Engineering, Indian Institute of Technology Indore, Indore-453446, INDIA*

[#]*Photonics Laboratory, Institute of Engineering & Technology, Devi Ahilya University, Khandwa Road, Indore-452001, INDIA*

Corresponding Author:
Shashi Prakash,
e-mail: sprakash_davv@rediffmail.com
tel: +91 9977186156
fax: +91 7312461116

The detection and analysis of defects in objects having complex shape is difficult using theoretical modeling. In present communication, we report our investigations undertaken towards detection of defects and non-destructive evaluation of low-modulus sample of Poly Methyl Metha Acrylate (PMMA) material using phase measuring deflectometry. Collimated light from He-Ne laser illuminates a sinusoidal grating. The resulting sinusoidal fringe pattern is projected onto the PMMA sample. The light reflected through the sample is recorded using Charge Couple Device (CCD) camera. When the sample was loaded along z-axis, the surface slope of the defective portions varies in an anomalous manner with respect to un-defective portions of the specimen. The prominent location and visualization of defects have been identified using windowed Fourier transform technique (WFT). The 3D phase plots reconstruct the profile of the specimen with presence of defects. The slope of the phase provides the information regarding defects and surface variation of the object. Experimental results conclusively establish the viability of the technique.

## Introduction

In recent years, due to technology advancement and interdisciplinary work culture, fabulous improvement has been made in field of non-destructive testing using optical techniques. Non-destructive testing comprises the test methods which are used to examine an object, material or system without impairing its future usefulness. Hence, it has been used for measurement of wide range of parameters in science and engineering. Various techniques have been reported for non-destructive testing. Major among them may be categorized as Optical testing, testing based on X-ray and Gamma ray, Acoustic waves, Electrical and electrostatic field variations, etc. Each of the technique has its area of influence within which it is most useful. Optical based techniques comprise of moiré methods [1], Holographic interferometry [2], speckle interferometry [3], grating shearing [4], and photo-elastic methods [5]. These have been widely used for non destructive testing.

The PMMA material has been investigated widely for various applications. Trivedi et al. proposed the use of Talbot interferometer for real time mapping of slope [6]. The work was extended towards testing the applicability of the technique for defect detection in bent plates. However, the method has a drawback that the distance between the two gratings is fixed and cannot be varied based on the demands of the set-up. Also, the wavefront reflected from the specimen tends to diverge and hence, the technique cannot be used for mapping surfaces with larger slope values. Tippur et al. proposed double grating lateral shearing interferometry called Coherent Gradient Sensing (CGS) and used it extensively in various engineering applications. The CGS method has been used to study the deformation near crack tip [7]. This technique was also used for simultaneous determination of slope and curvature from specularly reflective thin structures in real time [8]. However the technique requires a pair of grating for generation of fringe patterns.

In this communication, we report our investigations undertaken towards detection of defects and NDE of low modulus triangular shaped specimen of Poly Methyl Metha Acrylate (PMMA) material using deflectometry in conjunction with windowed Fourier fringe analysis. The experimental setup is simple and uses only a single grating. The accurate measurement of slope and detection of defect has been achieved by determining the phase plots of recorded fringe pattern. Only a single fringe pattern is required to extract the phase information and the phase plot. Defect detection has been successfully undertaken. Good agreement between theory and experiment is also observed.

## Methods

Schematic of the experimental arrangement for defect detection using deflectometric set-up has been shown in Fig. 1. Spatially filtered and collimated light from He-Ne laser illuminates the sinusoidal grating 'G' of period 0.25 mm. The sinusoidal fringe pattern is projected onto a specimen (triangular plate of low modulus PMMA material). Depending on the applied load onto specimen from the back surface, the reflected pattern is distorted. The reflected pattern is captured using CCD camera.

The recorded fringe pattern can be described as [9]

$$f(x,y) = a(x,y) + b(x,y)\cos[2\pi f_{ox}x + 2\pi f_{oy}y + \psi(x,y)] \qquad (1)$$

where $f(x,y)$ is the recorded intensity, $a(x,y)$ is the background intensity, $b(x,y)$ is the variation in the fringe visibility and $\psi(x,y)$ is the phase information of interest. The $f_{ox}$ and $f_{oy}$ are the spatial carrier frequency in the x and y direction, respectively. They are usually constant. It is assumed that the input fringe pattern $f(x,y)$ has already been discretized into pixels when it was captured by CCD camera, and thus *x* and *y* are integers. The WFT program developed in Matlab is used to extract, 3D phase maps from the recorded pattern. By applying the WFT and Inverse WFT as per the procedure given in [9], the filtered wrapped phase $\psi(x,y)$ can be extracted as:

$$\psi(x,y) = \text{angle}\,[\bar{f}(x,y)] \qquad (2)$$

where, $\bar{f}(x,y)$ denotes the filtered fringe pattern obtained after thresholding. The deflection angle $\Phi(x,y)$ can be written as,

$$\Phi(x,y) = \frac{p}{2\pi d}\psi(x,y); \qquad (3)$$

Where p is the period of the grating and d is distance between grating and the specimen. By applying phase unwrapping algorithm over $\psi(x,y)$, we obtain a continuous phase map. Unwrapped phase map has been obtained using the strategy followed by Kemao et al. [10]. The phase so determined is plotted against the pixel position. The deflection angle $\Phi(x,y)$ is related to $\psi(x,y)$ and is effectively the measure of the slope of the rays in the direction perpendicular to the grating lines.

When the test surface is without defects, a characteristic fringe pattern due to the slope of the surface is obtained. However, if there are an area of stress concentration, the slope around those regions varies in non-isocentric fashion, and hence, these areas of defects etc are prominently visualized. Thus, based on the change in the slope pattern, the nature of defects and their effect on the stress field of bent plate can easily be predicted.

## Experimental arrangement

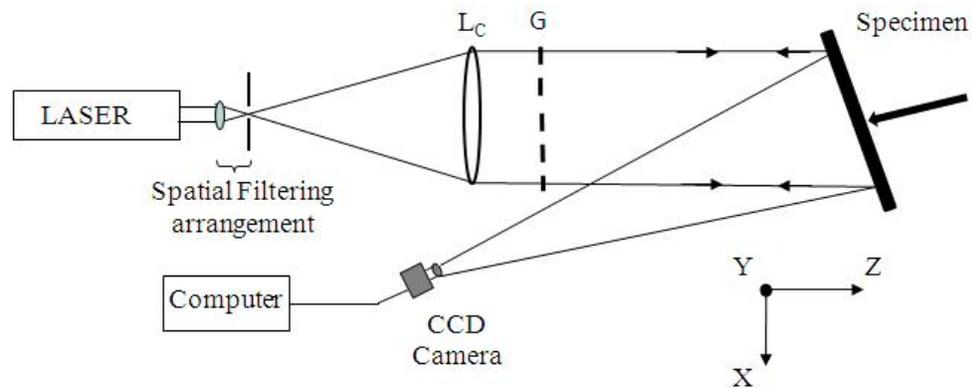

*Fig.1: Schematic of the experimental setup for detection of defect of PMMA plate.*

Fig. 1 shows the experimental setup for NDE and detecton of defect of PMMA plate. Light from He-Ne Laser of 15mW ($\lambda$=0.6328 µm) was used as a light source. The beam from He-Ne laser is spatially filtered using an arrangement comprising of microscope objective of magnification 60X and pinhole of 5 µm diameter. Collimating lens, of focal length 250mm, is used to produce a collimated laser beam. A translation stage of resolution 1 µm has been used to set the beam in collimation position. The collimated light is passing through sinusoidal grating and is projected onto the specimen. The reflected light from the specimen is directly recorded using CCD camera.

## Results

When a light beam is reflected from the object under study, the beam acquires a phase change, depending on the surface slope. The surface slope varies as the object was stressed by known amount. Low modulus specimen of PMMA material has been used during the experimentation. To analyze the defects, an equilateral triangular plate has been tested under flawless and defective mode. The equilateral-triangle-shaped plate has each side of 40mm and thickness of 2mm. It was mounted using a specially fabricated mechanical mount such that it was bound along the three corners. Artificial defects were created by making holes of approximately 1mm diameter and depth varying from 0.5mm to 1.5mm, at the back surface of the plate. Collimated light from the laser after passing through grating is incident onto the front surface of traingular plate. Under unstressed condition of the traingular plate, the back-reflected light beam remains collimated and parallel lines were observed onto the screen. When the specimen was loaded centrally along the negative z-direction by an amount of 0.3mm using a precision translating device, the slope fringes appeared on the screen.

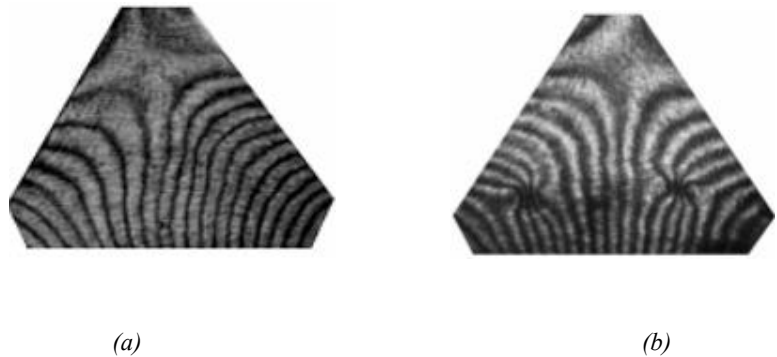

*(a)					(b)*

*Fig.2: Recorded fringe pattern for triangular plate (a) Flawless specimen loaded by 0.3mm (b) defective specimen loaded by 0.3mm.*

Fig. 2(a) corresponds to the slope contour along the horizontal direction for the non-defective plate. When the defective plate was clamped and centrally loaded, the surface slope of the defective portions varies in an anomalous manner with respect to un-defective portions of the specimen. The effect of defects was very prominent in terms of the change in orientation and spacing of the fringes at the defect site as shown in Fig. 2(b). Two defects were created on the plate (as shown in Fig. 2(b)) which are distingushed clearly in terms of change in orientation and shape of fringe patterns around the defect.

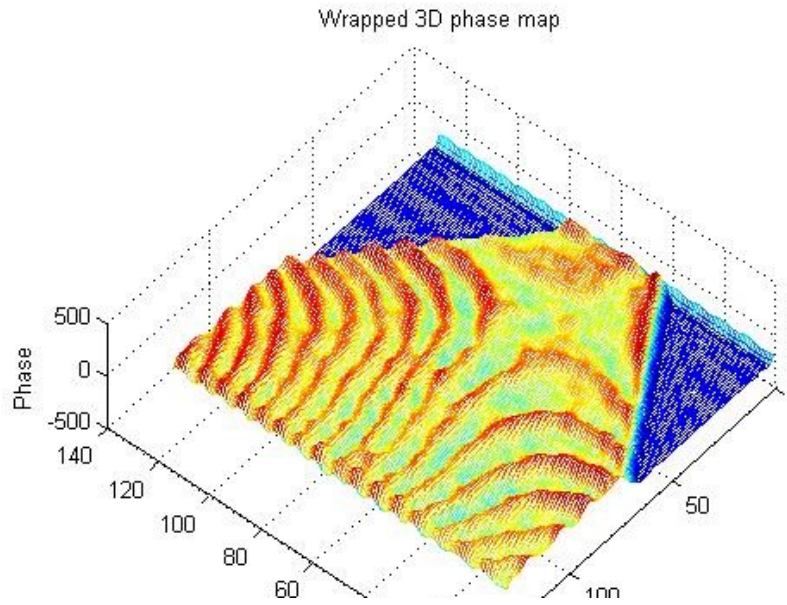

(a)

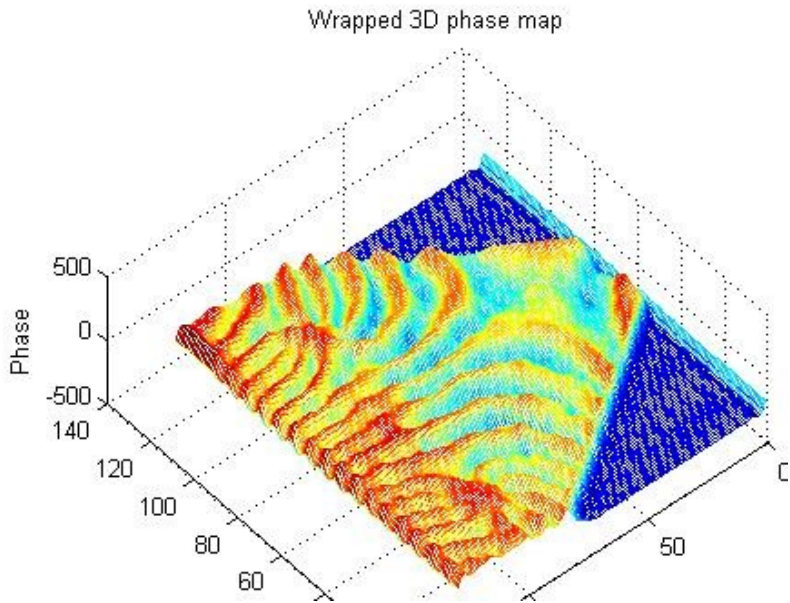

(b)

*Fig.3: 3D map for the triangular plate loaded along negative z-direction. (a) flawless triangular plate loaded by 0.3mm (b) defected triangular plate loaded by 0.3mm.*

To evaluate the phase values at each point of the interferogram a computer program based on the WTM algorithm [9] has been developed. In WFT algorothms, the fringe pattern is divided into a number of local fringe regions with a moving window. The choice of the window depends on the requirement; we used the Gaussian window in this work. The Fourier transform spectrum of each local fringe pattern is obtained using the moving window. All the local spectra are superimposed to obtain the complete spectra of the fringe pattern. In the next step, the fundamental spectrum of the whole fringe pattern's spectrum is extracted. Taking the inverse FT, the phase information of the fringe pattern is retrieved. Undertaking WFT by appropriate filtering scheme and thresholding, the wrapped phase has been determined.

To obtain reliable phase map, the field corresponding to the wrapped phase map is scanned and $2\pi$ is added or subtracted every time an edge is detected. The phase values so obtained are plotted against the pixel values using MATLAB toolbox. Fig. 3(a) and Fig. 3(b) shows three dimensional plot of the evaluated phase map with respect to the pixel position along x-axis for flawless and defective traingular plate, respectively. The colour bar has been used to depict the local variation in the defective portions and slope values.

In case of centrally loaded traingular plate bounded along its circumference, as per the thin plate theory, slope pattern is represented by [11]

$$\frac{\partial w}{\partial x} = \frac{4x w_{max}}{a^2} \log\left(\frac{\sqrt{x^2 + y^2}}{a}\right) \quad (4)$$

where, $w_{max}$ denotes the maximum out-of-plane displacement along negative z-direction, a denotes plate radius and (x, y) denote the coordinates of the traingular plate. There is a good agreement between theoretical slope and measured slope. An averaged error of 2.2% was obtained with respect to the theoretical prediction.

## Conclusion

In conclusion, we have successfully demonstrated the application of windowed Fourier fringe analysis technique for measurement of slope and defect detection of bent plates. Ttriangular shaped PMMA plates have been tested using the technique. Based on the variation in slope values for regions where defects are located with respect to the flawless regions region, the defects are prominently visualized. Improved accuracy in the measurement of slope and defect detection was obtained using automated fringe analysis technique.

## Acknowledgement

This publication is an outcome of the R&D work undertaken project under the Visvesvaraya PhD Scheme of Ministry of Electronics & Information Technology, Government of India, being implemented by Digital India Corporation.